\documentclass[letter]{aa} 

\usepackage{graphics}
\usepackage{longtable}
\usepackage{txfonts}
\usepackage{graphicx}

\begin{document}

\title{The environment of the fast rotating star Achernar\thanks{Based on observations made with ESO Telescopes at Paranal Observatory under programs 078.D-0295(C), (D) and (E).}}
\subtitle{II. Thermal infrared interferometry with VLTI/MIDI}
\titlerunning{Thermal infrared interferometry of Achernar with VLTI/MIDI}
\authorrunning{P. Kervella et al.}
\author{
P. Kervella\inst{1}
\and
A. Domiciano~de~Souza\inst{2}
\and
S. Kanaan\inst{2}
\and
A. Meilland\inst{3}
\and
A. Spang\inst{2}
\and
Ph. Stee\inst{2}
}
\offprints{P. Kervella}
\mail{Pierre.Kervella@obspm.fr}
\institute{
LESIA, Observatoire de Paris, CNRS UMR 8109, UPMC, Univ. Paris Diderot, 5 place Jules Janssen,
92195 Meudon Cedex, France
\and
Lab. H. Fizeau, CNRS UMR 6525, Univ. de Nice-Sophia Antipolis, Observatoire de la C\^ote dÕAzur, 06108 Nice Cedex 2, France
\and
Max-Planck-Intitut f\"ur Radioastronomie, Auf dem Hugel 69, 53121 Bonn, Germany
}
\date{Received ; Accepted}
\abstract
{As is the case of several other Be stars, Achernar is surrounded by an envelope, recently detected by near-IR interferometry.}
{We search for the signature of circumstellar emission at distances of a few stellar radii from Achernar, in the thermal IR domain.}
{We obtained interferometric observations on three VLTI baselines in the $N$ band (8-13\,$\mu$m), using the MIDI instrument.}
{ From the measured visibilities, we derive the angular extension and flux contribution of the $N$ band circumstellar emission in the polar direction of Achernar. The interferometrically resolved polar envelope contributes $13.4 \pm 2.5\,\%$ of the photospheric flux in the $N$ band, with a full width at half maximum of $9.9 \pm 2.3$\,mas ($\approx 6\,R_\star$). This flux contribution is in good agreement with the photometric IR excess of 10-20\% measured by fitting the spectral energy distribution. Due to our limited azimuth coverage, we can only establish an upper limit of 5-10\,\% for the equatorial envelope. We compare the observed properties of the envelope with an existing model of this star computed with the SIMECA code.}
{ The observed extended emission in the thermal IR along the polar direction of Achernar is well reproduced by the existing SIMECA model. Already detected at 2.2\,$\mu$m, this polar envelope is most probably an observational signature of the fast wind ejected by the hot polar caps of the star.}

\maketitle

\section{Introduction}

The southern Be star \object{Achernar} ($\alpha$\,Eri, \object{HD 10144}) has received much interest since its strongly distorted photosphere was resolved by long-baseline interferometry (Domiciano de Souza et al.~\cite{domiciano03}), with major and minor axes of respectively $\theta = 2.13 \pm 0.05$ and $1.51 \pm 0.02$\,milliarcseconds (Kervella \& Domiciano de Souza~\cite{kervella06}, hereafter K06).
Due to its extremely fast rotation ($v \sin i \approx 250$\,km.s$^{-1}$) and consequent flattening, the von Zeipel effect (von Zeipel~\cite{vonzeipel24}) causes the polar caps to be overheated: the polar effective temperature could be higher than 20\,000\,K, compared to $\lesssim 10\,000$\,K at the equator (Jackson et al.~\cite{jackson04}; Kanaan et al.~Ê\cite{kanaan08}, hereafter Ka08). The high radiative pressure at the poles creates a fast polar wind that was detected in the near infrared by K06, where its flux reaches $4.7 \pm 0.3$\,\% of the photosphere. In addition to this circumstellar envelope (hereafter CSE), Kervella \& Domiciano de Souza~(\cite{kervella07}) discovered a close-in companion of Achernar, of spectral type A1V-A3V (Kervella et al.~\cite{kervella08}).
In the present Letter, we report new interferometric observations of \object{Achernar} in the thermal infrared domain, using the VLTI/MIDI instrument. After a description of our measurements (Sect.~\ref{observations}), we derive the contribution and typical angular scale of the polar CSE of Achernar using a simple Gaussian model and compare them to SIMECA model predictions (Sect.~\ref{simplemodel}).

\section{Observations\label{observations}}

\subsection{Instrumental setup and data processing\label{instrument}}

MIDI (Leinert et al.~\cite{leinert03}; Ratzka et al.~\cite{ratzka07}) is the mid-infrared two-telescope beam combiner of the Very Large Telescope Interferometer (VLTI; Glindemann et al. \cite{glindemann04}). This instrument is a classical Michelson interferometer working in the astronomical $N$ band (7.6--13.3\,$\mu$m).
For the reported observations of Achernar, we used a prism with a spectral resolution of $R=\lambda/\Delta \lambda \simeq 30$ to obtain spectrally dispersed fringes.
During the observations, the secondary mirrors of the two Unit Telescopes were chopping with a frequency of 2\,Hz to properly sample the sky background. 
\object{Achernar} was observed in 2006 and 2007, using three 8.2\,m telescope baselines (UT1-UT4, UT1-UT2, and UT3-UT4), and the {\tt SCIPHOT} mode of MIDI. In this observing mode, the photometry of each telescope is recorded simultaneously with the interferometric signals, allowing a more accurate calibration of the visibilities. 
The average dates of each observation, with the corresponding projected baseline length $B$ and position angle $PA$ are given in Table~\ref{midi_log}.
Each star or calibrator observation corresponds to a time on target recording interferometric fringes of 3\,min, followed by approximately 5\,min of photometric calibrations.
For the raw data processing, we used two different software packages: {\tt MIA} developed at the Max-Planck-Institut f\"ur Astronomie and {\tt EWS} developed at the Leiden Observatory ({\tt MIA+EWS}\footnote{http://www.strw.leidenuniv.nl/$\sim$nevec/MIDI/index.html}, version 1.5.2) in order to extract the calibrated squared visibilities $V^2(\lambda)$ (Chesneau~\cite{chesneau07}).
We found a good agreement between the results of the MIA and EWS packages within the error bars. In the following we will refer to the results obtained with the EWS package.
Our calibrator, $\delta$\,Phe (\object{HD 9362}, G9III), was chosen in the Cohen et al.~(\cite{cohen99}) catalogue of spectrophotometric standards for infrared wavelengths.
$\delta$\,Phe is located relatively close to \object{Achernar} on the sky ($8.2^\circ$), and is of comparable brightness in the $N$ band (9.5\,Jy at 12\,$\mu$m vs. $\approx 16$\,Jy for Achernar). It is almost unresolved by the interferometer in the $N$ band, with $\theta_{\rm LD} = 2.24 \pm 0.02$\,mas (Bord\'e et al.~Ê\cite{borde02}).
The calibrated squared visibilities of Achernar are listed in online Table~\ref{visib_table}.
As MIDI operates in the diffraction limited regime of the UTs, the effective field of view diameter is equal to $\approx 0.26\arcsec$ at $\lambda=10\,\mu$m, much larger than the angular size of \object{Achernar} and its CSE. An imaging campaign in the $N$ band (Kervella \& Domiciano de Souza~\cite{kervella07}) uncovered a faint companion (Achernar~B) located $\approx 0.3^{\prime\prime}$ away from the star. Considering that the flux contribution of B is only $\approx 2$\% of A in the $N$ band, we assume in the following that its impact on the MIDI visibilities is negligible compared to their accuracy, and refer to Achernar~A simply as ``Achernar".

\begin{table}
\caption{Log of the observations of Achernar and its calibrator.} 
\label{midi_log}
\begin{tabular}{llccrr}
\hline
\# & {\it Baseline} / Date & UTC & Target & $B$ (m) & $PA$ ($^\circ$)\\
\hline
\noalign{\smallskip}
 & \it UT1-UT4 \\
A & 2006-11-06 & 01:59:46 & $\alpha$\,Eri &  129.39 & 48.89 \\
B & 2006-11-06 & 02:26:23 & HD\,9362 & 129.66 & 54.55 \\
C & 2006-11-07 & 00:35:03 & $\alpha$\,Eri & 130.22 & 32.23 \\
D & 2006-11-07 & 01:05:39 & HD\,9362 & 129.93 & 40.06 \\
 & \it UT1-UT2 \\
E & 2006-11-06 & 00:32:53 & $\alpha$\,Eri & 52.38 & 4.94 \\
F & 2006-11-06 & 01:07:56 & HD\,9362 & 54.64 & 10.84 \\
G & 2006-11-06 & 03:15:57 & $\alpha$\,Eri & 49.58 & 29.69 \\
H & 2006-11-06 & 04:08:42 & HD\,9362 & 50.53 & 35.36 \\
 & \it UT3-UT4 \\
I & 2006-12-07 & 02:48:10 & $\alpha$\,Eri & 62.46 & 126.54 \\
J & 2006-12-07 & 03:11:45 & HD\,9362 & 61.76 & 132.33 \\
K$^{*}$ & 2007-06-28 & 07:15:29 & HD\,9362 & 39.28 & 50.04 \\
L$^{*}$ & 2007-06-28 & 07:51:38 & $\alpha$\,Eri & 46.10 & 52.25 \\
\hline
\end{tabular}
\begin{list}{}{}
\item[$^{*}$] Due to a photometric instability, \#K and \#L were rejected.
\end{list}
\end{table}

\subsection{Visibilities \label{visibsection}}

\begin{figure}[]
\begin{center}
\includegraphics[width=8.3cm]{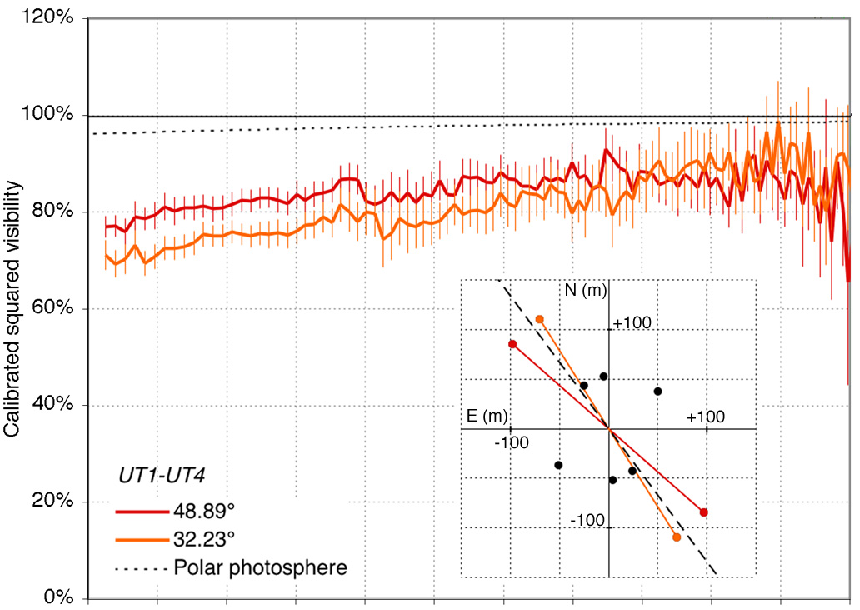}
\includegraphics[width=8.3cm]{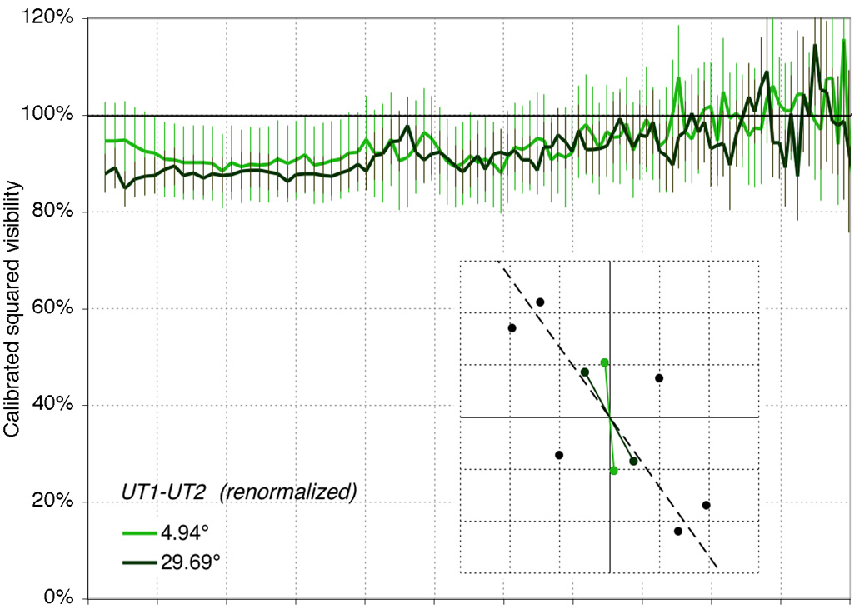}
\includegraphics[width=8.3cm]{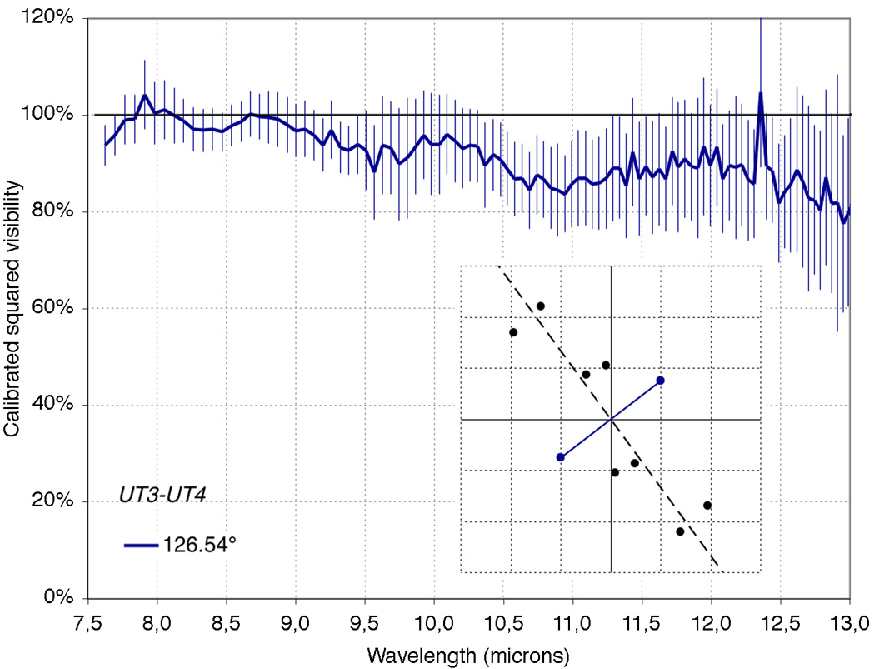}
\caption{Squared visibilities of Achernar. The error bars are the statistical dispersion of the $V^2$ measurements from the {\tt EWS} software. The position angles of each projected baseline are given in degrees. The inset diagrams show each baseline in the $(\lambda u,\lambda v)$ plane, and the polar direction of Achernar (dashed line). Thanks to the spectral coverage of MIDI, a relatively broad range of spatial frequencies is sampled simultaneously for each baseline (see Fig.~\ref{gaussianmodel}). The dotted line in the upper plot shows the polar $V^2$ function of Achernar, for $B = 130$\,m. The UT1-UT2 $V^2$ values were renormalized (see Sect.~\ref{visibsection} for details).}
\label{visibilities}
\end{center}
\end{figure}

The calibrated visibilities of \object{Achernar} are presented in Fig.~\ref{visibilities}.
One problem we encountered is that the $V^2$ spectra for our two UT1-UT2 baseline observations \#E and \#G originally reached ``unphysical" values larger than unity. Such a behavior was noticed by Chesneau~(\cite{chesneau07}), and is caused by an incorrect estimation of the $\kappa$ coefficients that characterize the internal photometric transmission of the instrument. The resulting visibility spectrum is affected by a multiplicative bias larger than unity, uniform with wavelength. We could identify the cause for this behavior in the data sets \#F and \#H. The derived interferometric transfer function $T^2$ from these calibrator observations is systematically lower than observation \#B of the same calibrator, that was made (in time) in between observations \#F and \#H. The ratio $\gamma(\#\mathrm{X})=T^2(\#\mathrm{X})/T^2(\#\mathrm{B})$ is respectively $\gamma(\#\mathrm{F}) = 0.86 \pm 0.07$ and $\gamma(\#\mathrm{H}) = 0.82 \pm 0.03$. As the seeing was stable over all our observations obtained on 2006-11-06 ($0.6-0.9\arcsec$ in the visible), such a variation of $T^2$ cannot be explained by changing atmospheric conditions. The instrumental cause for this bias on the calibrator $V^2$ spectrum is not identified, but could be linked to an instrumental polarization problem.
Our other observations are apparently free of such a ``$\kappa$-induced" bias. The fact that we obtain nearly the same $V^2$ spectrum for two different nights on the UT1-UT4 baseline (Fig.~\ref{visibilities}, top) gives confidence in their calibration. For this reason, we uniformly multiplied the $V^2$ spectra obtained on the UT1-UT2 baseline by the relevant $\gamma$ factors, whose uncertainties are reflected in the $V^2$ error bars (Fig.~\ref{visibilities}, middle).

Achernar shows a squared visibility deficit on the longer UT1-UT4 baseline (Fig.~\ref{visibilities}, top, $B \approx 130$\,m), that is almost aligned with the rotation axis of the star. Such a deficit ($\Delta V^2 \approx 20-30\%$) can be explained by the presence of a resolved CSE component, as discussed in Sect.~\ref{simplemodel}.
The visibilities on the UT3-UT4 baseline along the direction of the equator of the star show almost no resolution. However, the shorter projected baseline length of $\approx 62$\,m reduces the sensitivity of the interferometer to moderately extended emission.

\subsection{Spectrophotometry\label{spectrophot}}

\begin{figure}[]
\centering
\includegraphics[bb=15 10 365 145, width=8.7cm, angle=0]{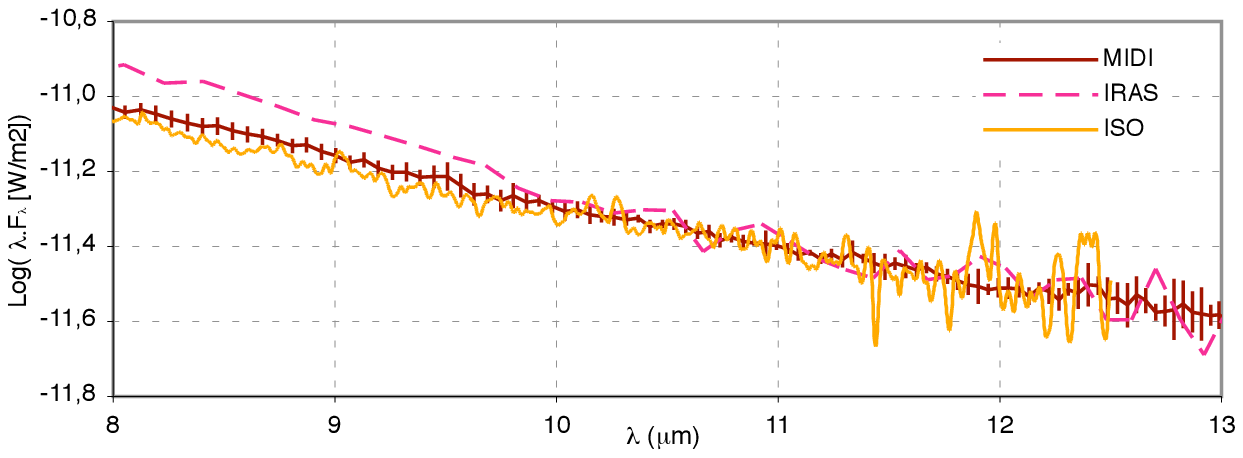}
\caption{Absolutely calibrated MIDI spectrum of Achernar, using HD\,9362 as a spectrophotometric standard star, with the spectra from IRAS LRS and ISO SWS (PWS processing) superposed.}
\label{spectro}
\end{figure}

The absolutely calibrated spectrum of \object{Achernar} presented in Fig.~\ref{spectro} was obtained by dividing the average MIDI spectrum by the average spectrum of its calibrator \object{HD 9362}, and then multiplying the result by the template spectrum from Cohen et al.~(\cite{cohen99}). The agreement with the IRAS spectrum (Volk \& Cohen~\cite{volk89}) is satisfactory, although an excess can be noticed in the IRAS data between 8 and 10\,$\mu$m, compared to the MIDI and ISO spectra. This could be attributed to a different activity level of the star for these two observations.
According to Fig.~14 of Ka08, our MIDI observations were obtained in a state of increasing H$\alpha$ emission of Achernar, after a minimum occuring around 2000-2002.
The IRAS observations took place in 1983, a year during which \object{Achernar} was in a decreasing, moderate activity state (Balona et al.~\cite{balona87}). The ISO SWS spectrum (Sloan et al.~\cite{sloan03}) was obtained in 1996, when the star was also in a decreasing activity state (Vinicius et al.~\cite{vinicius06}).

\begin{figure}[]
\centering
\includegraphics[bb=12 8 363 223, width=8.7cm, angle=0]{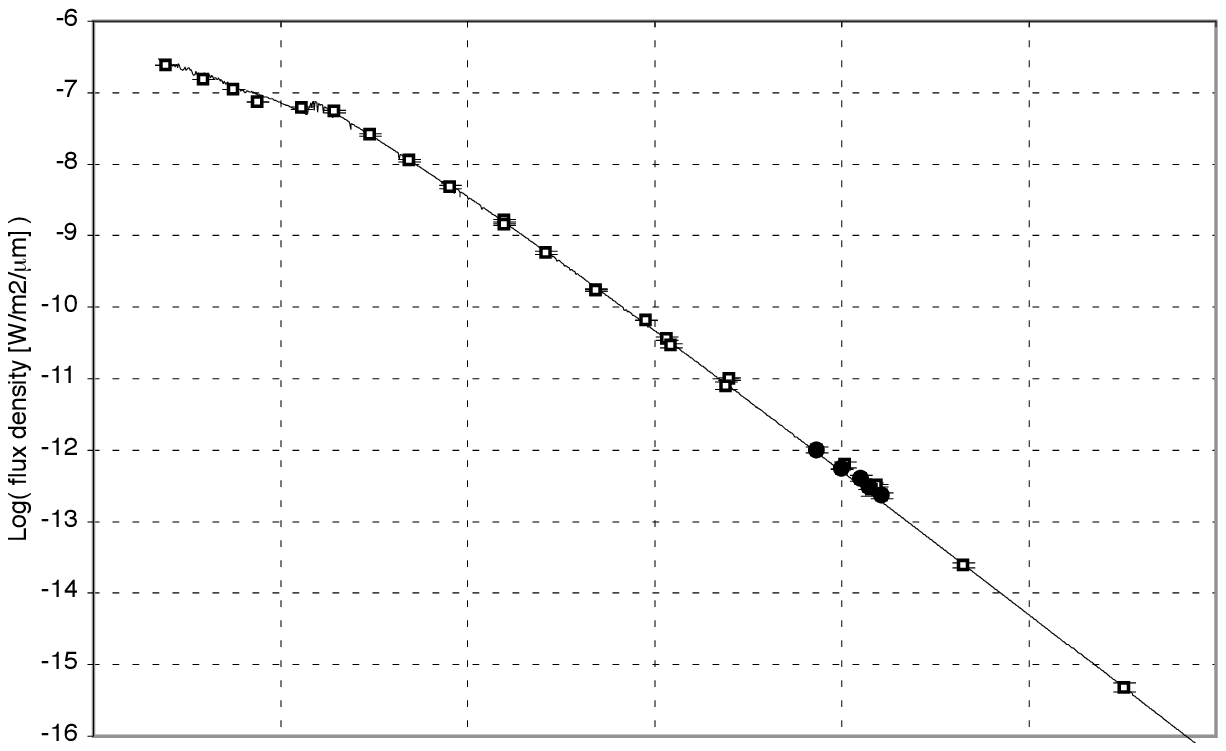}
\includegraphics[bb=12 8 363 143, width=8.8cm, angle=0]{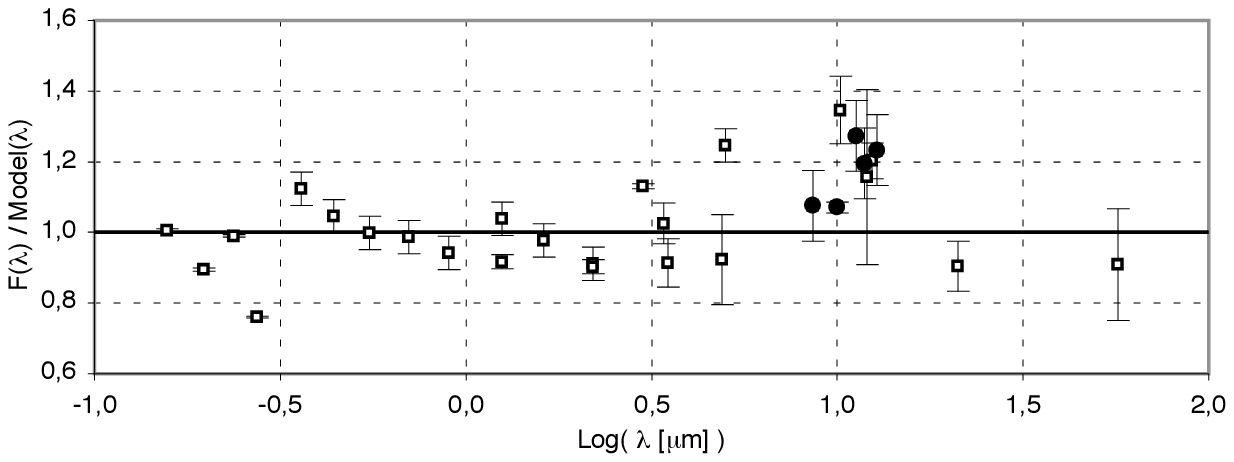}
\caption{Spectral energy distribution of Achernar. The open squares are measurements from the literature. The MIDI and VISIR points (Kervella \& Domiciano de Souza~\cite{kervella07}) are shown as solid dots.}
\label{sed}
\end{figure}

The spectral energy distribution (SED) model presented in Fig.~\ref{sed} was taken from the database of Castelli \& Kurucz~(\cite{castelli03}) using an average $T_{\rm eff}$ of 15\,000\,K and $\log g = 3.5$ (Levenhagen \& Leister~\cite{levenhagen06}; Lovekin et al.~\cite{lovekin06}), for solar metallicity.
The average angular diameter was set to $\theta_{\rm LD} = 1.79$\,mas in order to match the observed broadband photometry in the $V$ band, taken as a fiducial value. We chose this band as the contribution from the CSE is expected to be small in the visible. This value is very close to the arithmetic average of the polar and equatorial angular diameters (1.82\,mas) measured by K06. We are aware that this SED model is not physically realistic (it ignores in particular the von Zeipel effect), but Lovekin et al.~(\cite{lovekin06}) showed that the deviation from a rotating star SED is reasonably small.
The photometry was taken from Ducati~(\cite{ducati02}) for the $U$ to $N$ broadband photometry, Thompson et al.~(\cite{thompson78}) for the UV, ISO (Kessler et al.~\cite{kessler03}), IRAS (IPAC~\cite{ipac86}), and COBE/DIRBE (Smith et al.~\cite{smith04}) for the IR. We also used the VLT/VISIR photometry obtained by Kervella \& Domiciano de Souza~(\cite{kervella07}), as well as the average flux measured with MIDI between 8 and 12\,$\mu$m (Fig.~\ref{spectro}). As shown in Fig.~\ref{sed} (bottom), an excess emission of $\approx 10-20\%$ of the photospheric flux is present around $\lambda = 10\,\mu$m, but not at 25 and 60\,$\mu$m (IRAS measurements).

\section{Extended emission\label{simplemodel}}

\begin{figure*}[]
\centering
\includegraphics[bb=10 10 730 280, width=17cm, angle=0]{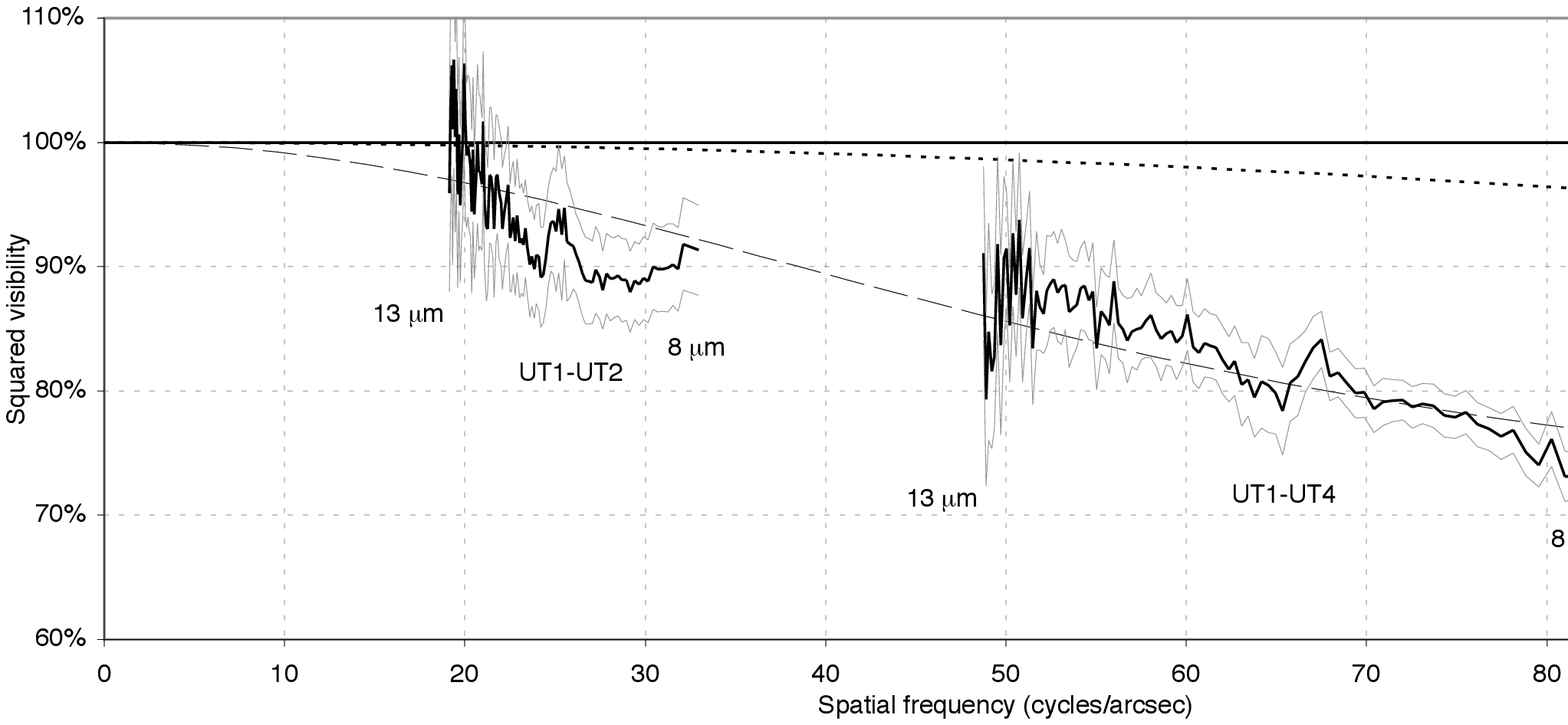}
\caption{Observed $V^2$ values of Achernar and adjusted photosphere+Gaussian envelope model (dashed curve) along the polar direction (separate averages of UT1-UT4 and UT1-UT2 baselines), as a function of spatial frequency. The polar photospheric $V^2$ is shown as a dotted curve.}
\label{gaussianmodel}
\end{figure*}

To estimate the relative flux contribution and spatial extension of the CSE of Achernar,  we use the simple model of a uniformly bright photosphere surrounded by a Gaussian CSE, with a full width at half maximum $\rho$ and a flux contribution relatively to the photosphere $\alpha = f_\mathrm{CSE}/f_\star$.
In this model, we assume that the CSE extension is independent of wavelength over the $N$ band. This type of model was already used by K06, and the interested reader is referred to these authors' Sect.~3.3 to 3.5 for details. In the present work, we restrict the fitting process to the polar direction of Achernar, i.e. to the UT1-UT4 and UT1-UT2 baselines, that are approximately aligned (Fig.~\ref{visibilities}). We fix the photospheric polar angular size to $\theta_{\rm pol}=1.51$\,mas (K06). The adjusted $V^2$ model is:
\begin{equation}
V_\mathrm{model}^2(\rho, \alpha, \theta_\mathrm{pol}, \nu) = \left( \frac{V_\star + \alpha V_\mathrm{CSE}}{1 + \alpha} \right)^2
\end{equation}
where  the photospheric $(V_\star)$ and CSE $(V_\mathrm{CSE})$ visibilities are:
\begin{equation}
V_\star(\theta_\mathrm{pol}, \nu) = \left| \frac{2\,J_1( \pi\, \theta_\mathrm{pol}\, \nu )}{\pi\, \theta_\mathrm{pol}\, \nu} \right|,\ \ V_\mathrm{CSE}(\rho, \nu) = \exp\left[ -\frac{\left( \pi\, \rho\, \nu \right)^2}{4\,\ln 2} \right],
\end{equation}
where $\rho$ is the FWHM of the CSE, and $\nu = B/\lambda$ the spatial frequency of the interferometric measurement.
The result of this fit is shown in Fig.~\ref{gaussianmodel}. The derived polar CSE parameters are a FWHM of $\rho = 9.9 \pm 2.3$\,mas and a flux contribution of $\alpha = 13.4 \pm 2.5$\,\% relatively to the photosphere, on average over the $N$ band. The reduced $\chi^2$ of the fit is satisfactory at 1.6. The measured CSE flux is comparable to the excess measured photometrically of 10-20\% (Sect.~\ref{spectrophot}).

Along the equatorial direction, the visibilities obtained on the UT3-UT4 baseline (Fig.~\ref{visibilities}, bottom) are only marginally different from unity. We can only set an upper limit of $\approx 5\%$ to the CSE flux contribution at $8\,\mu$m and $\approx 10\%$ at $12\,\mu$m in the equatorial direction of the star, at spatial scales larger than $20-30$\,mas. The possibility still exists that a significant equatorial CSE contribution exists in this direction, but it should then be smaller than $\approx 10\,R_\star$.
A comparison of the photometric infrared excess measured in Sect.~\ref{spectrophot} and the resolved polar CSE flux observed with MIDI also leaves little space for a bright equatorial component. We can roughly estimate the maximum equatorial CSE emission to $\approx 5-10\%$ at all spatial scales, in the $N$ band.

For the present discussion, we choose to check the agreement of Ka08's Achernar model with the parameters derived from our Gaussian CSE model fitting, rather than directly fit the SIMECA model to the MIDI data. The reason for this indirect approach is the limited coverage in baseline orientation and length of the MIDI data that make the convergence of the fit difficult.
The model of Achernar was computed by Ka08 using the SIMECA code (Stee \& Bittar~\cite{stee01}) for their analysis of VLTI/VINCI data. SIMECA has been developed to model the environment of active hot stars. It computes line profiles, spectral energy distributions (SEDs), and intensity maps in lines and the continuum, which can be directly compared to spectroscopic, photometric and high angular resolution observations. The best model of Ka08 for the epoch of VINCI observations (2002-2003) is that of a polar wind with an opening angle of about 20$\degr$ in the $K$ band (the model parameters are listed in Table~1 of Ka08). From this same SIMECA model, we derive in the $N$ band a half-maximum CSE radius of $6.8\,R_\star$, giving an angular FWHM of $\approx 10$\,mas considering the $\approx 7\,R_\odot$ polar radius of Achernar and its $\pi= 22.68 \pm 0.57$ parallax (ESA~\cite{esa97}). The associated $N$ band flux contribution of the CSE is 11\% of the photosphere.
The model extension and flux contribution are both in excellent agreement with the results of our MIDI observations (Sect.~\ref{simplemodel}).

The SIMECA model predicts CSE fluxes of $\approx 2\times$ and $\approx 11\times$ the photosphere at 25 and 60\,$\mu$m. Fig.~\ref{sed} shows no such excess in the IRAS photometry. This discrepancy may be due to the low activity of Achernar during IRAS observations, or more probably to inaccuracies of the model in this wavelength range.

\section{Conclusion}

From new interferometric observations in the thermal infrared domain ($8-13\,\mu$m), we resolve an extended CSE along the polar direction of \object{Achernar}, whose total flux is $13.4 \pm 2.5$\,\% of the photosphere, with a FWHM of $9.9 \pm 2.3$\,mas ($\approx 6\,R_\star$). This flux contribution is consistent with the photometric infrared excess of 10-20\% measured in the same wavelength domain, and with the predictions by the SIMECA model of Ka08. This convergence strengthens the plausibility of the presence of a fast polar wind ejected from the overheated polar caps of the star.
According to the scenario proposed by Ka08, \object{Achernar} is currently in a disk formation phase. Unfortunately, our limited coverage of the equatorial direction of Achernar with the MIDI data restricts our sensitivity. We could only set an upper flux limit of $\approx 10\%$ for an extended CSE in this direction. This non-detection is consistent with the K06 results in the $K$ band.

\begin{acknowledgements}
We thank Dr. O. Chesneau for his help in the processing of MIDI data, and the referee Dr. D. R. Gies for helpful comments on our manuscript. This research used the SIMBAD and VizieR databases at CDS, Strasbourg (France), and NASA's ADS bibliographic services. We used ISO (an ESA project) and NASA/IPAC Archive data.
We received the support of PHASE, the high angular resolution partnership between  ONERA, Observatoire de Paris, CNRS, and University Denis Diderot Paris 7.
\end{acknowledgements}

{}

\Online

\longtab{2}{
\begin{longtable}{rccccc}
\caption{MIDI squared visibilities of Achernar. The position angle is counted positively East of North ($N=0^\circ$, $E=90^\circ$). \label{visib_table}}\\
\hline \hline
Baseline & UT1-UT4   & UT1-UT4   & UT1-UT2   & UT1-UT2   & UT3-UT4   \\
Proj. $B$ (m) & 129.39   & 130.22   & 52.38   & 49.58   & 62.46   \\
Pos. Angle ($^\circ$) & 48.89   & 32.23   & 4.94   & 29.69   & 126.54   \\
\hline                     
\noalign{\smallskip}                     
$\lambda$\,($\mu$m) & $ V^2(\lambda) \pm \sigma $ & $ V^2(\lambda) \pm \sigma $ & $ V^2(\lambda) \pm \sigma $ & $ V^2(\lambda) \pm \sigma $ & $ V^2(\lambda) \pm \sigma $ \\
\hline \hline                    
\endfirsthead
\caption{continued.}\\
\hline \hline
\endhead
\hline
7.623 & $ 0.770 \pm 0.022 $ & $ 0.711 \pm 0.031 $ & $ 0.947 \pm 0.079 $ & $ 0.879 \pm 0.041 $ & $ 0.938 \pm 0.042 $ \\
7.695 & $ 0.771 \pm 0.022 $ & $ 0.693 \pm 0.028 $ & $ 0.946 \pm 0.079 $ & $ 0.890 \pm 0.042 $ & $ 0.959 \pm 0.042 $ \\
7.767 & $ 0.760 \pm 0.028 $ & $ 0.703 \pm 0.031 $ & $ 0.948 \pm 0.081 $ & $ 0.847 \pm 0.037 $ & $ 0.991 \pm 0.052 $ \\
7.839 & $ 0.791 \pm 0.032 $ & $ 0.732 \pm 0.030 $ & $ 0.936 \pm 0.079 $ & $ 0.867 \pm 0.036 $ & $ 0.992 \pm 0.051 $ \\
7.911 & $ 0.786 \pm 0.023 $ & $ 0.694 \pm 0.026 $ & $ 0.925 \pm 0.082 $ & $ 0.873 \pm 0.039 $ & $ 1.043 \pm 0.072 $ \\
7.982 & $ 0.793 \pm 0.028 $ & $ 0.708 \pm 0.026 $ & $ 0.921 \pm 0.077 $ & $ 0.875 \pm 0.038 $ & $ 1.005 \pm 0.065 $ \\
8.053 & $ 0.812 \pm 0.029 $ & $ 0.726 \pm 0.025 $ & $ 0.908 \pm 0.078 $ & $ 0.888 \pm 0.038 $ & $ 1.011 \pm 0.062 $ \\
8.123 & $ 0.802 \pm 0.028 $ & $ 0.725 \pm 0.024 $ & $ 0.906 \pm 0.076 $ & $ 0.894 \pm 0.039 $ & $ 0.999 \pm 0.058 $ \\
8.194 & $ 0.810 \pm 0.024 $ & $ 0.729 \pm 0.025 $ & $ 0.901 \pm 0.076 $ & $ 0.875 \pm 0.037 $ & $ 0.988 \pm 0.046 $ \\
8.263 & $ 0.809 \pm 0.025 $ & $ 0.737 \pm 0.025 $ & $ 0.901 \pm 0.077 $ & $ 0.881 \pm 0.037 $ & $ 0.971 \pm 0.045 $ \\
8.333 & $ 0.812 \pm 0.024 $ & $ 0.754 \pm 0.025 $ & $ 0.901 \pm 0.077 $ & $ 0.870 \pm 0.036 $ & $ 0.969 \pm 0.045 $ \\
8.402 & $ 0.807 \pm 0.026 $ & $ 0.751 \pm 0.023 $ & $ 0.899 \pm 0.077 $ & $ 0.878 \pm 0.034 $ & $ 0.971 \pm 0.044 $ \\
8.471 & $ 0.809 \pm 0.026 $ & $ 0.752 \pm 0.024 $ & $ 0.884 \pm 0.077 $ & $ 0.876 \pm 0.036 $ & $ 0.966 \pm 0.040 $ \\
8.539 & $ 0.817 \pm 0.024 $ & $ 0.760 \pm 0.024 $ & $ 0.902 \pm 0.077 $ & $ 0.877 \pm 0.036 $ & $ 0.979 \pm 0.044 $ \\
8.607 & $ 0.825 \pm 0.023 $ & $ 0.754 \pm 0.023 $ & $ 0.894 \pm 0.076 $ & $ 0.885 \pm 0.038 $ & $ 0.986 \pm 0.042 $ \\
8.674 & $ 0.824 \pm 0.024 $ & $ 0.750 \pm 0.023 $ & $ 0.899 \pm 0.076 $ & $ 0.887 \pm 0.036 $ & $ 1.003 \pm 0.047 $ \\
8.741 & $ 0.829 \pm 0.021 $ & $ 0.756 \pm 0.024 $ & $ 0.896 \pm 0.077 $ & $ 0.886 \pm 0.035 $ & $ 0.997 \pm 0.047 $ \\
8.807 & $ 0.830 \pm 0.021 $ & $ 0.754 \pm 0.028 $ & $ 0.898 \pm 0.079 $ & $ 0.883 \pm 0.038 $ & $ 0.996 \pm 0.054 $ \\
8.874 & $ 0.825 \pm 0.026 $ & $ 0.758 \pm 0.027 $ & $ 0.910 \pm 0.080 $ & $ 0.879 \pm 0.035 $ & $ 0.991 \pm 0.057 $ \\
8.939 & $ 0.818 \pm 0.025 $ & $ 0.752 \pm 0.028 $ & $ 0.899 \pm 0.080 $ & $ 0.863 \pm 0.034 $ & $ 0.980 \pm 0.058 $ \\
9.004 & $ 0.836 \pm 0.030 $ & $ 0.761 \pm 0.027 $ & $ 0.907 \pm 0.080 $ & $ 0.877 \pm 0.037 $ & $ 0.968 \pm 0.055 $ \\
9.069 & $ 0.824 \pm 0.028 $ & $ 0.772 \pm 0.028 $ & $ 0.917 \pm 0.080 $ & $ 0.878 \pm 0.039 $ & $ 0.972 \pm 0.058 $ \\
9.133 & $ 0.837 \pm 0.027 $ & $ 0.775 \pm 0.029 $ & $ 0.896 \pm 0.080 $ & $ 0.878 \pm 0.036 $ & $ 0.959 \pm 0.052 $ \\
9.197 & $ 0.840 \pm 0.025 $ & $ 0.790 \pm 0.032 $ & $ 0.901 \pm 0.082 $ & $ 0.876 \pm 0.038 $ & $ 0.938 \pm 0.055 $ \\
9.260 & $ 0.846 \pm 0.026 $ & $ 0.778 \pm 0.030 $ & $ 0.905 \pm 0.083 $ & $ 0.872 \pm 0.039 $ & $ 0.970 \pm 0.060 $ \\
9.322 & $ 0.867 \pm 0.030 $ & $ 0.816 \pm 0.035 $ & $ 0.910 \pm 0.082 $ & $ 0.880 \pm 0.038 $ & $ 0.934 \pm 0.054 $ \\
9.385 & $ 0.869 \pm 0.032 $ & $ 0.800 \pm 0.041 $ & $ 0.922 \pm 0.083 $ & $ 0.886 \pm 0.037 $ & $ 0.928 \pm 0.051 $ \\
9.446 & $ 0.867 \pm 0.031 $ & $ 0.780 \pm 0.040 $ & $ 0.924 \pm 0.081 $ & $ 0.897 \pm 0.034 $ & $ 0.940 \pm 0.060 $ \\
9.507 & $ 0.822 \pm 0.040 $ & $ 0.801 \pm 0.049 $ & $ 0.950 \pm 0.089 $ & $ 0.883 \pm 0.038 $ & $ 0.927 \pm 0.082 $ \\
9.568 & $ 0.816 \pm 0.040 $ & $ 0.798 \pm 0.049 $ & $ 0.921 \pm 0.090 $ & $ 0.915 \pm 0.049 $ & $ 0.882 \pm 0.098 $ \\
9.628 & $ 0.823 \pm 0.044 $ & $ 0.744 \pm 0.059 $ & $ 0.921 \pm 0.102 $ & $ 0.919 \pm 0.054 $ & $ 0.938 \pm 0.100 $ \\
9.688 & $ 0.841 \pm 0.042 $ & $ 0.756 \pm 0.055 $ & $ 0.950 \pm 0.096 $ & $ 0.944 \pm 0.046 $ & $ 0.932 \pm 0.096 $ \\
9.747 & $ 0.821 \pm 0.051 $ & $ 0.788 \pm 0.056 $ & $ 0.905 \pm 0.101 $ & $ 0.947 \pm 0.062 $ & $ 0.899 \pm 0.116 $ \\
9.806 & $ 0.840 \pm 0.047 $ & $ 0.775 \pm 0.060 $ & $ 0.913 \pm 0.104 $ & $ 0.979 \pm 0.059 $ & $ 0.911 \pm 0.125 $ \\
9.864 & $ 0.819 \pm 0.044 $ & $ 0.770 \pm 0.045 $ & $ 0.934 \pm 0.097 $ & $ 0.923 \pm 0.056 $ & $ 0.935 \pm 0.099 $ \\
9.922 & $ 0.838 \pm 0.036 $ & $ 0.780 \pm 0.051 $ & $ 0.965 \pm 0.091 $ & $ 0.908 \pm 0.049 $ & $ 0.960 \pm 0.092 $ \\
9.980 & $ 0.834 \pm 0.051 $ & $ 0.777 \pm 0.045 $ & $ 0.952 \pm 0.088 $ & $ 0.920 \pm 0.044 $ & $ 0.940 \pm 0.106 $ \\
10.037 & $ 0.866 \pm 0.041 $ & $ 0.782 \pm 0.038 $ & $ 0.926 \pm 0.091 $ & $ 0.923 \pm 0.044 $ & $ 0.940 \pm 0.103 $ \\
10.093 & $ 0.836 \pm 0.032 $ & $ 0.799 \pm 0.045 $ & $ 0.904 \pm 0.089 $ & $ 0.910 \pm 0.046 $ & $ 0.961 \pm 0.083 $ \\
10.149 & $ 0.834 \pm 0.035 $ & $ 0.816 \pm 0.042 $ & $ 0.893 \pm 0.084 $ & $ 0.894 \pm 0.043 $ & $ 0.946 \pm 0.086 $ \\
10.205 & $ 0.875 \pm 0.036 $ & $ 0.795 \pm 0.039 $ & $ 0.900 \pm 0.085 $ & $ 0.883 \pm 0.045 $ & $ 0.931 \pm 0.083 $ \\
10.260 & $ 0.871 \pm 0.040 $ & $ 0.802 \pm 0.033 $ & $ 0.917 \pm 0.089 $ & $ 0.900 \pm 0.049 $ & $ 0.938 \pm 0.075 $ \\
10.315 & $ 0.873 \pm 0.033 $ & $ 0.804 \pm 0.048 $ & $ 0.903 \pm 0.087 $ & $ 0.917 \pm 0.044 $ & $ 0.936 \pm 0.075 $ \\
10.369 & $ 0.862 \pm 0.037 $ & $ 0.800 \pm 0.046 $ & $ 0.909 \pm 0.089 $ & $ 0.888 \pm 0.044 $ & $ 0.897 \pm 0.088 $ \\
10.423 & $ 0.860 \pm 0.035 $ & $ 0.810 \pm 0.049 $ & $ 0.897 \pm 0.088 $ & $ 0.920 \pm 0.045 $ & $ 0.918 \pm 0.074 $ \\
10.477 & $ 0.883 \pm 0.035 $ & $ 0.840 \pm 0.050 $ & $ 0.880 \pm 0.084 $ & $ 0.923 \pm 0.049 $ & $ 0.908 \pm 0.077 $ \\
10.530 & $ 0.870 \pm 0.029 $ & $ 0.818 \pm 0.048 $ & $ 0.913 \pm 0.096 $ & $ 0.918 \pm 0.042 $ & $ 0.890 \pm 0.076 $ \\
10.582 & $ 0.868 \pm 0.038 $ & $ 0.810 \pm 0.049 $ & $ 0.935 \pm 0.091 $ & $ 0.927 \pm 0.044 $ & $ 0.869 \pm 0.077 $ \\
10.635 & $ 0.853 \pm 0.035 $ & $ 0.843 \pm 0.051 $ & $ 0.928 \pm 0.091 $ & $ 0.907 \pm 0.047 $ & $ 0.869 \pm 0.071 $ \\
10.687 & $ 0.854 \pm 0.028 $ & $ 0.838 \pm 0.059 $ & $ 0.938 \pm 0.093 $ & $ 0.908 \pm 0.050 $ & $ 0.845 \pm 0.079 $ \\
10.739 & $ 0.847 \pm 0.051 $ & $ 0.836 \pm 0.045 $ & $ 0.952 \pm 0.094 $ & $ 0.886 \pm 0.050 $ & $ 0.877 \pm 0.082 $ \\
10.790 & $ 0.872 \pm 0.040 $ & $ 0.827 \pm 0.044 $ & $ 0.948 \pm 0.095 $ & $ 0.934 \pm 0.052 $ & $ 0.868 \pm 0.079 $ \\
10.841 & $ 0.865 \pm 0.051 $ & $ 0.857 \pm 0.046 $ & $ 0.907 \pm 0.088 $ & $ 0.934 \pm 0.050 $ & $ 0.850 \pm 0.084 $ \\
10.892 & $ 0.870 \pm 0.038 $ & $ 0.843 \pm 0.045 $ & $ 0.920 \pm 0.082 $ & $ 0.959 \pm 0.055 $ & $ 0.845 \pm 0.097 $ \\
10.942 & $ 0.862 \pm 0.029 $ & $ 0.839 \pm 0.052 $ & $ 0.913 \pm 0.104 $ & $ 0.944 \pm 0.055 $ & $ 0.838 \pm 0.079 $ \\
10.993 & $ 0.903 \pm 0.040 $ & $ 0.797 \pm 0.058 $ & $ 0.921 \pm 0.096 $ & $ 0.926 \pm 0.049 $ & $ 0.858 \pm 0.089 $ \\
11.042 & $ 0.870 \pm 0.035 $ & $ 0.825 \pm 0.046 $ & $ 0.965 \pm 0.095 $ & $ 0.967 \pm 0.054 $ & $ 0.871 \pm 0.099 $ \\
11.092 & $ 0.876 \pm 0.042 $ & $ 0.805 \pm 0.057 $ & $ 0.980 \pm 0.104 $ & $ 0.928 \pm 0.047 $ & $ 0.870 \pm 0.098 $ \\
11.141 & $ 0.849 \pm 0.033 $ & $ 0.851 \pm 0.048 $ & $ 0.959 \pm 0.099 $ & $ 0.928 \pm 0.051 $ & $ 0.858 \pm 0.093 $ \\
11.190 & $ 0.854 \pm 0.034 $ & $ 0.855 \pm 0.053 $ & $ 0.930 \pm 0.094 $ & $ 0.931 \pm 0.058 $ & $ 0.859 \pm 0.096 $ \\
11.239 & $ 0.931 \pm 0.042 $ & $ 0.845 \pm 0.055 $ & $ 0.967 \pm 0.096 $ & $ 0.936 \pm 0.061 $ & $ 0.869 \pm 0.104 $ \\
11.288 & $ 0.912 \pm 0.048 $ & $ 0.793 \pm 0.066 $ & $ 0.954 \pm 0.094 $ & $ 0.964 \pm 0.064 $ & $ 0.891 \pm 0.110 $ \\
11.336 & $ 0.892 \pm 0.044 $ & $ 0.827 \pm 0.055 $ & $ 0.955 \pm 0.098 $ & $ 0.993 \pm 0.054 $ & $ 0.891 \pm 0.106 $ \\
11.384 & $ 0.885 \pm 0.042 $ & $ 0.843 \pm 0.062 $ & $ 0.982 \pm 0.097 $ & $ 0.960 \pm 0.060 $ & $ 0.853 \pm 0.108 $ \\
11.432 & $ 0.842 \pm 0.040 $ & $ 0.826 \pm 0.060 $ & $ 0.927 \pm 0.099 $ & $ 0.934 \pm 0.067 $ & $ 0.924 \pm 0.112 $ \\
11.480 & $ 0.885 \pm 0.056 $ & $ 0.875 \pm 0.055 $ & $ 0.959 \pm 0.093 $ & $ 0.960 \pm 0.062 $ & $ 0.869 \pm 0.118 $ \\
11.528 & $ 0.882 \pm 0.036 $ & $ 0.863 \pm 0.065 $ & $ 0.991 \pm 0.109 $ & $ 0.955 \pm 0.064 $ & $ 0.895 \pm 0.103 $ \\
11.575 & $ 0.880 \pm 0.047 $ & $ 0.889 \pm 0.058 $ & $ 0.962 \pm 0.095 $ & $ 0.985 \pm 0.066 $ & $ 0.872 \pm 0.118 $ \\
11.622 & $ 0.858 \pm 0.038 $ & $ 0.907 \pm 0.068 $ & $ 0.936 \pm 0.097 $ & $ 0.925 \pm 0.058 $ & $ 0.888 \pm 0.116 $ \\
11.669 & $ 0.863 \pm 0.049 $ & $ 0.876 \pm 0.049 $ & $ 0.948 \pm 0.109 $ & $ 0.916 \pm 0.071 $ & $ 0.868 \pm 0.108 $ \\
11.716 & $ 0.870 \pm 0.050 $ & $ 0.863 \pm 0.060 $ & $ 1.004 \pm 0.110 $ & $ 0.897 \pm 0.058 $ & $ 0.926 \pm 0.112 $ \\
11.763 & $ 0.851 \pm 0.059 $ & $ 0.876 \pm 0.070 $ & $ 1.078 \pm 0.107 $ & $ 0.956 \pm 0.067 $ & $ 0.893 \pm 0.134 $ \\
11.809 & $ 0.877 \pm 0.041 $ & $ 0.893 \pm 0.073 $ & $ 0.966 \pm 0.105 $ & $ 0.967 \pm 0.062 $ & $ 0.910 \pm 0.122 $ \\
11.856 & $ 0.867 \pm 0.060 $ & $ 0.901 \pm 0.074 $ & $ 0.951 \pm 0.091 $ & $ 1.003 \pm 0.081 $ & $ 0.894 \pm 0.111 $ \\
11.902 & $ 0.851 \pm 0.051 $ & $ 0.907 \pm 0.062 $ & $ 0.987 \pm 0.109 $ & $ 0.965 \pm 0.077 $ & $ 0.890 \pm 0.145 $ \\
11.948 & $ 0.894 \pm 0.044 $ & $ 0.885 \pm 0.079 $ & $ 1.012 \pm 0.097 $ & $ 0.986 \pm 0.085 $ & $ 0.936 \pm 0.143 $ \\
11.994 & $ 0.851 \pm 0.057 $ & $ 0.922 \pm 0.051 $ & $ 1.018 \pm 0.092 $ & $ 0.929 \pm 0.078 $ & $ 0.896 \pm 0.126 $ \\
12.039 & $ 0.871 \pm 0.057 $ & $ 0.890 \pm 0.072 $ & $ 0.947 \pm 0.094 $ & $ 0.938 \pm 0.062 $ & $ 0.934 \pm 0.121 $ \\
12.085 & $ 0.838 \pm 0.043 $ & $ 0.886 \pm 0.047 $ & $ 1.047 \pm 0.111 $ & $ 0.941 \pm 0.068 $ & $ 0.869 \pm 0.112 $ \\
12.130 & $ 0.810 \pm 0.043 $ & $ 0.925 \pm 0.059 $ & $ 0.993 \pm 0.094 $ & $ 0.896 \pm 0.092 $ & $ 0.896 \pm 0.113 $ \\
12.175 & $ 0.901 \pm 0.070 $ & $ 0.860 \pm 0.065 $ & $ 1.003 \pm 0.105 $ & $ 0.961 \pm 0.069 $ & $ 0.892 \pm 0.147 $ \\
12.220 & $ 0.823 \pm 0.067 $ & $ 0.845 \pm 0.060 $ & $ 0.988 \pm 0.096 $ & $ 0.992 \pm 0.080 $ & $ 0.898 \pm 0.125 $ \\
12.265 & $ 0.903 \pm 0.059 $ & $ 0.928 \pm 0.071 $ & $ 0.957 \pm 0.102 $ & $ 1.038 \pm 0.069 $ & $ 0.866 \pm 0.125 $ \\
12.310 & $ 0.919 \pm 0.060 $ & $ 0.862 \pm 0.065 $ & $ 0.973 \pm 0.123 $ & $ 1.007 \pm 0.066 $ & $ 0.858 \pm 0.112 $ \\
12.355 & $ 0.844 \pm 0.068 $ & $ 0.872 \pm 0.077 $ & $ 0.972 \pm 0.115 $ & $ 1.061 \pm 0.100 $ & $ 1.048 \pm 0.155 $ \\
12.399 & $ 0.907 \pm 0.082 $ & $ 0.967 \pm 0.073 $ & $ 1.035 \pm 0.094 $ & $ 1.092 \pm 0.111 $ & $ 0.895 \pm 0.110 $ \\
12.443 & $ 0.877 \pm 0.072 $ & $ 0.878 \pm 0.069 $ & $ 1.061 \pm 0.140 $ & $ 0.942 \pm 0.071 $ & $ 0.886 \pm 0.109 $ \\
12.487 & $ 0.865 \pm 0.082 $ & $ 0.989 \pm 0.083 $ & $ 1.021 \pm 0.098 $ & $ 0.942 \pm 0.074 $ & $ 0.818 \pm 0.123 $ \\
12.530 & $ 0.827 \pm 0.057 $ & $ 0.879 \pm 0.078 $ & $ 1.008 \pm 0.100 $ & $ 0.890 \pm 0.079 $ & $ 0.842 \pm 0.117 $ \\
12.574 & $ 0.886 \pm 0.058 $ & $ 0.943 \pm 0.076 $ & $ 1.009 \pm 0.102 $ & $ 1.004 \pm 0.078 $ & $ 0.855 \pm 0.138 $ \\
12.617 & $ 0.877 \pm 0.094 $ & $ 0.938 \pm 0.091 $ & $ 1.043 \pm 0.131 $ & $ 0.873 \pm 0.091 $ & $ 0.887 \pm 0.171 $ \\
12.660 & $ 0.816 \pm 0.129 $ & $ 0.858 \pm 0.088 $ & $ 1.042 \pm 0.113 $ & $ 1.044 \pm 0.120 $ & $ 0.863 \pm 0.178 $ \\
12.702 & $ 0.871 \pm 0.109 $ & $ 0.965 \pm 0.092 $ & $ 1.013 \pm 0.116 $ & $ 0.996 \pm 0.102 $ & $ 0.828 \pm 0.192 $ \\
12.745 & $ 0.841 \pm 0.088 $ & $ 0.815 \pm 0.077 $ & $ 0.987 \pm 0.121 $ & $ 1.146 \pm 0.127 $ & $ 0.825 \pm 0.155 $ \\
12.787 & $ 0.777 \pm 0.111 $ & $ 0.855 \pm 0.075 $ & $ 0.969 \pm 0.146 $ & $ 1.052 \pm 0.164 $ & $ 0.804 \pm 0.164 $ \\
12.828 & $ 0.893 \pm 0.141 $ & $ 0.803 \pm 0.111 $ & $ 1.075 \pm 0.113 $ & $ 1.049 \pm 0.146 $ & $ 0.872 \pm 0.182 $ \\
12.869 & $ 0.739 \pm 0.120 $ & $ 0.848 \pm 0.085 $ & $ 1.078 \pm 0.130 $ & $ 0.988 \pm 0.109 $ & $ 0.818 \pm 0.187 $ \\
12.910 & $ 0.905 \pm 0.099 $ & $ 0.917 \pm 0.100 $ & $ 0.940 \pm 0.124 $ & $ 0.979 \pm 0.102 $ & $ 0.818 \pm 0.266 $ \\
12.950 & $ 0.806 \pm 0.117 $ & $ 0.923 \pm 0.083 $ & $ 1.157 \pm 0.151 $ & $ 0.988 \pm 0.163 $ & $ 0.775 \pm 0.184 $ \\
12.990 & $ 0.655 \pm 0.213 $ & $ 0.892 \pm 0.130 $ & $ 0.944 \pm 0.123 $ & $ 0.926 \pm 0.168 $ & $ 0.799 \pm 0.195 $ \\
13.030 & $ 0.783 \pm 0.175 $ & $ 0.813 \pm 0.087 $ & $ 0.808 \pm 0.150 $ & $ 0.851 \pm 0.140 $ & $ 0.823 \pm 0.209 $ \\
\end{longtable}
}

\end{document}